\documentclass{svproc}
\usepackage{url}
\usepackage{comment}

\usepackage{a4wide}
\usepackage{amsmath,amssymb, amsfonts, thm-restate,mathtools}
\usepackage{comment}
\usepackage{hyperref}
\usepackage{enumerate}
\usepackage{enumitem}
\usepackage{xparse}
\usepackage{tikz}
\usepackage{bbm}
\usepackage{cite}
\usepackage{pgf} 
\usepackage{graphicx}
\usepackage{euscript}
\usepackage{multirow}
\usepackage{soul}  

\usepackage{algorithm}
\usepackage[noend]{algpseudocode}

\usepackage{rotating}

\usetikzlibrary{calc}
\usetikzlibrary{decorations.markings}
\usetikzlibrary{arrows.meta, shapes.misc, positioning,patterns,snakes}
\usepackage{microtype}
\usepackage{color,colortbl}
 \definecolor{darkgrey}{gray}{0.35}
\definecolor{grey}{gray}{0.86}
\definecolor{lightgrey}{gray}{0.91}
\definecolor{ballblue}{rgb}{0, 0.5,0.5}
\definecolor{lightballblue}{rgb}{0, 0.8,0.8}
\definecolor{dbblue}{rgb}{0, 0.4,0.4}
\usepackage{caption}
\newcommand{\N}{\mathbb{N}}

\newcommand{\R}{\mathbb{R}}

\DeclareMathOperator*{\argmin}{arg\,min}

\newcommand{\US}[1]{\todo[color=green,inline,caption={}]{Ulysse: #1}}

\begin{document}
\mainmatter              
\title{Expressivity of Geometric Inhomogeneous Random Graphs - Metric and Non-Metric}
\titlerunning{Expressivity of GIRGs}  
%
\author{Benjamin Dayan\inst{1} \and Marc Kaufmann\inst{2}
 \and Ulysse Schaller\inst{3}}
\authorrunning{Dayan, Kaufmann, and Schaller} 
%
\tocauthor{Benjamin Dayan, Marc Kaufmann, Ulysse Schaller}
\institute{Institut für Theoretische Informatik, ETH Zürich, Zürich, Switzerland \\
\email{bdayan@ethz.ch},
\and
Institut für Theoretische Informatik, ETH Zürich, Zürich, Switzerland
\\ \email{marc.kaufmann@inf.ethz.ch},
\and
Institut für Theoretische Informatik, ETH Zürich, Zürich, Switzerland
\\ \email{ulysse.schaller@inf.ethz.ch}
}

\maketitle              

\begin{abstract}

Recently there has been increased interest in fitting generative graph models to real-world networks. In particular, Bläsius et al.\ have proposed a framework for systematic evaluation of the expressivity of random graph models. We extend this framework to Geometric Inhomogeneous Random Graphs (GIRGs). This includes a family of graphs induced by non-metric distance functions which allow capturing more complex models of partial similarity between nodes as a basis of connection - as well as homogeneous and non-homogeneous feature spaces. As part of the extension, we develop schemes for estimating the multiplicative constant and the long-range parameter in the connection probability. Moreover, we devise an algorithm for sampling Minimum-Component-Distance GIRGs whose runtime is linear both in the number of vertices and in the dimension of the underlying geometric space.
Our results provide evidence that GIRGs are more realistic candidates with respect to various graph features such as closeness centrality, betweenness centrality, local clustering coefficient, and graph effective diameter, while they face difficulties to replicate higher variance and more extreme values of graph statistics observed in real-world networks.

\end{abstract}

\section{Introduction}

Many complex, large-scale datasets contain vast amounts of relational data, which can be leveraged to model the dataset as a network. For some, like social, biological, or technological networks, their connectivity information itself is often at the core of any data analysis.
On the other hand, real-world data tends to be noisy, incomplete, or should not be directly processed due to privacy or data integrity concerns.
Generative graph models offer a simple method for quickly generating synthetic network data, preserving much of the structure of the real-world dataset of interest.
On the synthetic data, one can then perform experiments such as performance analysis of algorithms or investigate stochastic processes in the network. In the best case, suitable embeddings can alleviate severe congestion problems such as in internet routing~\cite{bog:papa:kriouk}. The external validity of insights on randomly generated graphs for practical instances depends however on the embedding quality~\cite{blaes:fried:katz:force}. Conducting research in this direction, Bläsius and Fischbeck found that the performance in the idealized setting translates surprisingly well to real-world networks~\cite{blaes:fisch}. In particular, they find that if increasing some parameter in the network models, for example locality, improves performance, then we should also expect better performance for real-world networks with high locality. Many real-world networks across various domains exhibit similar global properties\cite{vander}:
\begin{itemize}
    \item Scale-free degree distribution: there is no typical (bounded) range for the node degrees - they occur "at all scales". The most prominent example of this is the Pareto or power-law distribution.
    \item Sparsity: the number of connections in the network scales linearly with the number of nodes.
    \item Clustering: two neighbors of a given node tend to be connected to each other as well, despite the edge sparsity. This is often seen as an indication of the presence of geometry.
    \item Small-world property: polylogarithmically bounded average distance and diameter.
\end{itemize}

While graph models usually capture at least one of these properties well - Chung-Lu graphs exhibit a scale-free degree distribution, as do preferential attachment graphs, Random Geometric Graphs display clustering - they struggle or outright fail to capture all structural aspects simultaneously.
Chung-Lu graphs as well as Erdős-Rényi graphs have vanishing clustering coefficient, while Random Geometric Graphs have homogeneous degrees and large distances.
A promising family of versatile sparse graph models are Geometric Inhomogeneous Random Graphs (GIRGs)~\cite{bring:keusch:leng:girgs}, which combine the clustering of Random Geometric Graphs with the power-law degree distribution of Chung-Lu graphs.
It has further been shown that the choice of underlying distance function in a GIRG can modulate the robustness of the graph - measured in the size of separators, that is, edge sets whose deletion splits the giant component of the graph into two linearly-sized pieces~\cite{lengler:todorovic}. 
Since it seems difficult to seamlessly represent large sets of networks perfectly with one model, two approaches have been taken: either, one assesses models regarding how well they represent larger collections of networks \emph{``on average"} (to find a good one-size-fits-all model); or, one chooses a specific model for a given network based on its most salient features\cite{motal:aliak:habib}.
In the latter case, the first step is to determine what these features actually are \cite{aliak:motall:rashid:habib:mova, att:aliak, nagy:molontay}. Further examples of work in this direction aim to detect the presence of geometry in the graph \cite{liu:mohanty:schramm:yang, kriouk:clust, bub:ding:eldan:racz}. If this is answered in the positive, the next question is then to figure out the dimensionality of the underlying geometry\cite{alm:bog}.
This approach assumes implicitly that the parametric family of models considered is the best - or at least suitable - to describe the real-world network. For GIRGs, recently Friedrich et al.\ have found a linear-time algorithm which determines the dimension of the underlying space, leveraging the exponential decay of the clustering coefficient in the dimension~\cite{fried:goeb:stat}. Under this model prior, the examined networks exhibit low dimensionality (between 1 and 20 dimensions).

In the present work, we follow the one-size-fits-all approach. Building on the work of~\cite{blaes:fried:katz:krohm:strie}, we extend the proposed framework to include GIRGs. In their experiments (cf. 3.3. in \cite{blaes:fried:katz:krohm:strie}), the authors observed that the most difficult networks to model were social networks, in particular Facebook graphs. In our experiments, we focus on this most difficult subset of real-world networks, which makes up 104 out of the 219 networks analyzed in~\cite{blaes:fried:katz:krohm:strie}, with the hope that it is a set of graphs with consistent structure.

Another key finding was the observation that hyperbolic random graphs (which turn out to be a specific instance of GIRGs~\cite{bring:keusch:leng:girgs}) were the most realistic, outperforming the Chung-Lu, Erdös-Rényi and Barabási-Albert graphs. 

\paragraph{Our contribution}
We extend the framework introduced by Bläsius et al.\ in~\cite{blaes:fried:katz:krohm:strie} to cover GIRGs of dimension $d$ between 1 and 7, including a range of different underlying non-metric distance functions. In doing so, we provide techniques for estimating the leading constant $c$ and the long-range parameter $\alpha$ jointly in the connection probability~\eqref{eq:girg-connection}.
We further show how the power-law parameter can be estimated from the degree distribution. For the class of Minimum-Component-Distance GIRGs, we devise an $O(dn)$ sampling algorithm, which improves upon the exponential dimensional dependence of the linear-time sampling algorithm for metric GIRGs~\cite{bring:keusch:leng:girgs}. 
We also extend the latter algorithm to enable sampling GIRGs equipped with the cube topology through a simple coupling trick. We believe that these techniques will be useful to any practitioner who is considering the deployment of GIRGs to model real-world network data.
Our experiments provide evidence that GIRGs are more realistic candidates for various graph features, including closeness centrality, betweenness centrality, local clustering coefficient, and graph effective diameter. 

\section{Framework for Evaluating Generative Graph Models}\label{sec:framework}

In 2018, Bläsius et al.\ introduced a framework with the goal of serving as a systematic experimental testbed for evaluating the ability of generative graph models to realistically synthesize any given graph dataset~\cite{blaes:fried:katz:krohm:strie}. They had several questions in mind:
\begin{itemize}
    \item Which characteristics of real-world networks are captured by a specific model, which are not?
    \item Do certain models fit specific types of real-world networks, i.e. technological or social, better than others?
    \item Given two (asymptotically) similar models, do they generate graphs with similar properties in practice? One example here are graph models with the small-world property, such as Chung-Lu, preferential-attachment graphs, and GIRGs, in particular where the theoretically derived bounds are not always known to be tight.
\end{itemize}
Their approach to address these questions can be summarized as follows: 
\begin{enumerate}
    \item Select a set of generative graph models.
    \item Select a collection of real-world networks.
    
    \item For each generative graph model and each network, estimate the parameters of the model based on the network statistics.
    \item Generate a set of synthetic graphs using the estimated model parameters.
    \item Select a set of graph features and feed it into a classifier that aims to distinguish real-world networks from synthetic networks of a particular generative model.
\end{enumerate}
The authors' intention was that their modular approach could be adapted by including further generative graph models, additional parameters, and new real-world networks.
The used machine learning technique (classifier) should also be interchangeable.
As a proof-of-concept, they investigated the following generative graph models: Erdös-Rényi, Barabási-Albert, Chung-Lu, and hyperbolic random graphs. These were evaluated on 219 real-world networks retrieved from the Network Repository~\cite{rossi:ahmed}.
In order to make the description of their approach self-contained, we succinctly outline their methodology.

\textbf{Network collections.} First, choose a collection of real-world networks $C=\{G_1,...,G_c\}$ with $c=|C|$. Based on each model $M$ - using its generative mechanism - and each graph $G_i \in C$, generate an artificial graph $G_i^M$ which aims to mimic $G_i$ as closely as possible. The exact fitting procedure depends on the model but tends to be governed by choosing the model parameters to match a small set of empirical graph statistics exhibited by the real networks. When the input parameters are known to influence mainly one statistic of the generated graph and the relationship is monotone, the parameters can be estimated with a binary search. The collection of graphs generated using $M$ is denoted by $C^M=\{G_1^M,...,G_c^M\}$.

\textbf{Network features and feature cleaning.}
Next, choose a set of relevant graph statistics as a feature set $F = \{\Phi_1, ..., \Phi_f \}$, which collectively define a feature vector of a graph, $F(G) = (\Phi_1(G), ..., \Phi_f(G))$. For an entire collection of graphs $C$, by slight abuse of notation, $F(C)$ denotes the collection of feature vectors. Note that the selection of features here crucially influences how well the structural properties which distinguish different network types can be represented. In the discussion of our experimental findings, we will comment on the expressivity of the chosen baseline features and possible augmentations, in particular for GIRGs. All the features are computed on the largest connected component of the graphs.
Meaningless features are eliminated through numerical cleaning (deletion of all features which are undefined or infinite for at least one of the networks), variation cleaning (elimination of features with little predictive value because they assume similar values on most networks) and correlation grouping.
More precisely, for \emph{variation cleaning}, features are eliminated based on a normalized coefficient of variation \footnote{For a given feature, let $X$ be the vector containing the $c$ values it assumes in different graphs. Let $\sigma$ and $\mu$ denote the standard deviation and mean. Then the feature's normalized coefficient of variation is defined as $\frac{\sigma(X)}{\mu(X) \sqrt{c-1}}$} below a threshold of $1\%$. \emph{Correlation grouping} groups highly correlated features (having multiple very similar features does not add any predictive value).
For each group of correlated features, only the feature with the clearest semantics (given by a manually predefined order) of the group is used. The grouping proceeds by constructing a graph with features as nodes and edges connecting two features by an edge if their absolute Spearman's rank correlation coefficient is above $99\%$. Note that grouped features may have a smaller correlation than the threshold of $99\%$, as having correlation larger than some fixed threshold is not transitive, however being in the same connected component is.

\textbf{Distinguishing real and synthetic networks.}
Finally, we determine which pairs of collections can be distinguished based on which features. The queries answered are as follows. The input consists of a subset $F$ of all features and of two collections of graphs, usually the collection $C$ of real networks plus one collection $C^M$ for some model $M$. We then want to know how well $F(C)$ can be distinguished from $F(C^M)$. The classification task is as follows. The input for the classifier consists of a feature matrix $X \in \mathbb{R}^{2c \times f}$ and a binary vector $Y \in \{0,1\}^{2c}$ which classifies the features as belonging to $C$ (denoted by $0$) or to $C^M$ (denoted by $1$).
The binary classification model is then just a function $M: \mathbb{R}^{f} \to \{0,1\}$.
The predictions are then evaluated in terms of their \emph{accuracy}, that is the ratio of correctly classified examples.
We use $l-$fold cross validation, that is we split the data into $l$ random subsets of equal size which generate $l$ learned models. For each model, a single subset is used as the test dataset, and all other subsets are used as the training data.
More precisely, we use \emph{stratified} cross validation which ensures that the number of examples is the same for both classes. The total accuracy of the cross-validation is then defined as the arithmetic mean of the accuracies of all models. The classifiers used are support vector machines (SVMs) with the Gaussian radial basis function (rbf) kernel. The parameters for the SVM and the rbf kernel are selected by cross-validation using a grid search over the parameter space. The model with the best average test accuracy is used as the final model. All features used in the SVM are normalized to have zero mean and unit variance in the training data. The test set is scaled using the same parameters.

\textbf{Graph properties.}
The graph properties 
used can be divided into two categories. \emph{Single-value features} assign a single numerical value to a graph. The most basic among them include the number of nodes, the number of edges, the diameter (maximum length of a shortest path between any two nodes in the graph), and the effective diameter (upper bound on the shortest path lengths between $90\%$ of all node pairs).
The other category of features is distribution over nodes.
These features assign a value to each node, leading to distributions over all nodes in the graph. For each of these distributions we consider the arithmetic mean, the median, the first and third quartile, and the standard deviation. The simplest such measure is the degree distribution. The \emph{local clustering coefficient} of a vertex $v$ is the probability that two randomly selected neighbors of $v$ are connected. The $k$-core of a graph is obtained by successively removing all nodes with degree less than $k$. This leads to the measure of \emph{core centrality}, where each node is assigned the largest $k$ such that it is contained in the $k$-core. The \emph{betweenness centrality} measures for each vertex $v$ how many shortest paths between pairs of other nodes go through $v$, and the \emph{closeness centrality} of a node denotes its average distance to every other node in the graph. The \emph{Katz centrality} measures the importance of a node by its number of neighbors and the distance of all other nodes to these neighbors.
Distinguishing real and synthetic networks using the higher number of data points from distribution level features is too easy, and it is hence often hard to compare between generative graph models. Thus for distribution type features we just use the mean value as input to the classifier.

\section{Graph Models}\label{sec:graph-models}
We will consider undirected graphs with vertex set $V\coloneqq[n]$ and edge set denoted by $E$.

\textbf{Erdős-Rényi Graphs.}
As a classical baseline, we include the Erdős-Rényi random graph in our framework, similarly to~\cite{blaes:fried:katz:krohm:strie}. To generate the graph, each pair of vertices is connected by an edge independently of the others with some probability $p$. Thus the expected number of edges $p \cdot \frac{n(n-1)}{2}$ can be chosen. Given a real network, $n$ is set to the number of vertices and the connection probability is set to $p=\frac{2m}{n(n-1)}$, where $m$ is the number of edges in the network. 

\textbf{Barabási-Albert Graphs.}
Here, a graph is generated from an initial fixed graph of size $k$, to which $n-k$ nodes are added sequentially, each connecting to $k$ already present nodes with probability proportional to their (current) degree. Since $2k$ is the expected average degree in the graph, we derive $k$ from the empirical average degree, and $n$ is simply set to the empirical number of vertices. In~\cite{blaes:fried:katz:krohm:strie} two initial graphs were used, namely a complete graph and a cycle. However, the results are very similar, so here we restrict to complete graph initializations.

\textbf{Hyperbolic Random Graphs.} 
In this graph model, $n$ nodes draw their positions randomly in a disk within the hyperbolic plane, and pairs of vertices are then connected with a probability decreasing in their hyperbolic distance (see Definition 4.1 in~\cite{bring:keusch:leng:girgs} for a precise definition). The connection probability also depends on $n$ and the target average degree. The degree distribution then follows a power law whose exponent $\beta$ can be specified as an input. The influence of the hyperbolic distance is softened by a temperature parameter $T \in [0,1)$, which corresponds roughly to the inverse of the $\alpha$ parameter in the GIRG model. The number of vertices $n$ is then fit by estimating the number of isolated nodes because the largest connected component of a generated graph is typically smaller than $n$. The desired average degree is set to the empirical average degree. The power-law exponent $\beta$ is estimated based on the cumulative degree distribution. Finally, $T$ is fit by a binary search, using as a proxy the clustering coefficient which it influences - comparing its values in the synthetic graph and the real network.

\textbf{Chung-Lu Graphs.}
In Chung-Lu graphs~\cite{chung:lu}, each vertex $v\in V$ draws a weight $w_v$ independently from a power-law distribution with exponent $\tau$, and each pair of vertices $\{u,v\}$ is then connected independently with probability $\min\{cw_uw_v/W, 1\}$, where $W:=\sum_{v\in V} w_v$ is the sum of all weights and $c>0$ is a constant. This results in a graph where the expected degree of a vertex is proportional to its weight, and in particular the degree distribution follows a power-law with the same exponent $\tau$. We take $\tau\in(2,3)$, which ensures both that the graph is sparse (i.e.\ $|E|=\Theta(n)$) and has a (unique) giant component with high probability.

\subsection{Geometric Inhomogeneous Random Graphs}

We extend the experiments in~\cite{blaes:fried:katz:krohm:strie} to include Geometric Inhomogeneous Random Graphs (GIRGs) and variants thereof as graph models. GIRGs were introduced in~\cite{bring:keusch:leng:girgs} and combine the degree inhomogeneity of Chung-Lu graphs with an underlying geometric space. The vertices are assigned both a weight and a position in a given ground space. We will take the $d$-dimensional unit hypercube $[-\tfrac{1}{2},\tfrac{1}{2}]^d$ as ground space, equipped with either the torus topology or the usual Euclidean subspace topology. As distance functions we will use the classical max-norm $\|\cdot\|_\infty$ as well as various non-metric distances - obtained by taking a sequence of minima and maxima of component-wise distances - which we call Boolean distance functions and define below (note that this definition includes the max-norm). 

\begin{definition}\label{def:boolean-distance-function}
Let $d\in\N$ and consider the hypercube $[-\tfrac{1}{2},\tfrac{1}{2}]^d$ equipped with either the torus topology or the usual Euclidean subspace topology. A function $\|\cdot\| : [-\tfrac{1}{2},\tfrac{1}{2}]^d \rightarrow \R_{\geq 0}$ is a \emph{Boolean distance function} if it can be recursively defined as follows:
\begin{itemize}
    \item For $d = 1$, $\|x\| = |x|$.
    \item For $d > 1$, there exists a nonempty proper subset $S \subset \{1, \ldots, d\}$ of coordinates such that either $\|x\| = \max(\|(x_i)_{i \in S}\|_S, \|(x_i)_{i \notin S})\|_{\overline{S}})$ or $\|x\| = \min(\|(x_i)_{i \in S}\|_S, \|(x_i)_{i \notin S})\|_{\overline{S}})$, where $\|\cdot\|_S : [-\tfrac{1}{2},\tfrac{1}{2}]^{|S|} \rightarrow \R_{\geq 0}$ and $\|\cdot\|_{\overline{S}} : [-\tfrac{1}{2},\tfrac{1}{2}]^{d-|S|} \rightarrow \R_{\geq 0}$ are Boolean distance functions.
\end{itemize}
\end{definition}

\begin{definition}[GIRG]\label{def:girg}
    Let $d\in\N$ and consider the hypercube $[-\tfrac{1}{2},\tfrac{1}{2}]^d$ equipped with either the torus topology or the usual Euclidean subspace topology and a Boolean distance function $\|\cdot\| : [-\tfrac{1}{2},\tfrac{1}{2}]^d \rightarrow \R_{\geq 0}$. For parameters $\tau\in(2,3)$, $\alpha>1$ and $c>0$, a \emph{Geometric Inhomogeneous Random Graph} is obtained by the following three-step procedure:
    \begin{enumerate}[label=(\arabic*), leftmargin=1cm]
        \item Every vertex $v\in V$ draws i.i.d\ a weight $w_v$ from a power-law distribution with exponent $\tau$.

        \item Every vertex $v\in V$ draws independently and u.a.r.\ a position $x_v$ in $[-\tfrac{1}{2},\tfrac{1}{2}]^d$.

        \item Connect each pair of vertices $\{u,v\}$ independently with probability
        \begin{align}\label{eq:girg-connection}
            p_{uv} \coloneqq \min\Big\{c\Big(\frac{w_u w_v / W}{Vol_{\|\cdot\|}(\|x_u-x_v\|)}\Big)^{\alpha}, 1\Big\},
        \end{align}
        where $Vol_{\|\cdot\|}(r)$ denotes the volume (or Lebesgue measure) of a ball of radius $r$ in $[-\tfrac{1}{2},\tfrac{1}{2}]^d$ with respect to the chosen Boolean distance function $\|\cdot\|$, and $W:=\sum_{v\in V} w_v$ is the sum of all weights.
    \end{enumerate}
\end{definition}

In our experiments, we will consider GIRGs with different underlying Boolean distance functions $\|\cdot\|$ and/or of different dimensions $d$ as different models. Each of these models therefore has three parameters that need to be estimated: the power-law exponent $\tau$, which influences the tail of the degree distribution; the long-range parameter $\alpha$, which controls the number of long edges between small-weight vertices; and the constant $c$, which directly impacts the average degree.

After GIRGs were introduced in~\cite{bring:keusch:leng:girgs}, several of their theoretical properties were established, focusing on the torus topology and the geometry induced by the max-norm $\|\cdot\|_\infty$. The expected degree of a vertex is proportional to its weight, so the degree distribution of a GIRG follows a power-law with exponent $\tau$, which also implies that $|E|=\Theta(n)$. GIRGs have a high clustering coefficient (i.e\ $\Theta(1)$) and a unique linear-size connected component, which has a polylogarithmic diameter and a doubly logarithmic average distance. These properties (scale-free degree distribution, sparsity, local communities, small world phenomenon) have been observed empirically in a wide range of real-world networks, including social networks but also biological and technological networks. This makes the GIRG model an ideal candidate to study real-world networks theoretically or via computer simulations. On top of that, GIRGs can be generated quickly using a linear time algorithm described in~\cite{bring:keusch:leng:girgs} and implemented in~\cite{blaes:fried:katz:mey:pen:weya}.

\textbf{Cube vs. Torus.}
From a theoretician's point of view the distinction of \emph{T-GIRGs} (GIRGs with an underlying torus topology) and \emph{C-GIRGs} (GIRGs with the usual Euclidean topology of the hypercube) is not crucial, since the two models have the same asymptotic properties as the number of nodes grows to infinity. However, from a practitioner's point of view this distinction is relevant. Indeed, depending on the type of real-world network studied, the underlying topology could be toroidal (e.g.\ transport networks at the planet scale) or cubic (e.g.\ social networks in a given country, where geographic proximity has a huge influence on the friendship links between people). Moreover, the topology has a clear influence on some properties of the graph at a fixed size. For example, the diameter of a network will be smaller in a T-GIRG compared to a C-GIRG, since edges that wrap around the hypercube's boundary are present in the former but not in the latter~\cite{blaes:fisch}.

\textbf{Non-metric GIRGs.} 
Since it is a priori unclear for many networks if they are metric in nature (i.e.\ if the triangle inequality holds for them), an evaluation of realistic distance functions should include also non-metric distances such as the Minimum-Component-Distance as well as more complicated disjunctive normal forms of the coordinate distances (see Definition~\ref{def:boolean-distance-function}). Of particular interest are so-called outer-min distances which consist of taking the minimum of arbitrary combinations of distance functions. Imagine for instance a social network where each node is equipped with features that encode kinship, various aspects of employment, and membership in one or several social clubs. It is plausible to hypothesize that at least some of these features - such as sharing a parent - guarantee an edge between two nodes independently of all other features. Others may not suffice to produce an edge by themselves, but if two or three among them each yield a small component-wise distance, this again makes the existence of an edge very likely, independently of all other features.   For instance, two individuals may have the same (multinational) employer \emph{and} a workplace located in the same city - making it plausible that the two individuals work in the same branch office and thus making it very likely that they know each other - irrespective of their familial relation. Such cases are convenient to model with outer-min distances. Despite their relevance, only little is known theoretically about such non-metric distance functions. Lengler and Todorovic showed for instance that GIRGs equipped with the Minimum-Component-Distance (MCD-GIRGs) are much more robust than their metric counterparts in that they do not contain any sublinear separators – sublinear sets of edges which partition the giant component into linearly-sized disconnected vertex sets\cite{lengler:todorovic}. In our experiments we further consider a range of outer-min distances GIRGs for dimension $d\in\{1,2,3,4,5\}$.

\section{GIRG Sampling and Parameter Estimation}

In this section, we describe techniques for fitting the parameters of the GIRG model: the leading constant $c$ in the connection probability, the long-range parameter $\alpha$, as well as the power-law exponent $\tau$ of the weight distribution.
Note that for the final parameter of the model, namely the dimension $d$, we take an orthogonal approach and evaluate the quality of fit for a range of dimensions, treating $d$ as a parameter to be \emph{chosen}.

\subsection{Power-Law Parameter}

As shown in~\cite{bring:keusch:leng:avgd}, we have $\mathbb{E}[\mathrm{deg}(v)] = \Theta(w_v)$ in GIRGs for all vertices $v\in V$. Hence a GIRG's weights give a proxy for its degrees, which is why sampling those weights from a power-law distribution replicates the power-law degree distribution commonly observed in real-world networks. The empirical degree distribution of a graph at $k \in \mathbb{N}$ is defined as $\frac{|\{v \in V : \mathrm{deg}(v) = k\}|}{|V|}$. This discrete distribution's tail can be approximated with statistical estimators of the power-law exponent $\tau$, and the weights are then generated from a continuous power-law distribution with that exponent. We use the python package \verb+powerlaw+, which finds a threshold $x_{\min}$ where the power-law tail begins, and fits $\tau$ in the region $[x_{\min},\infty)$ using maximum likelihood estimation. The optimal $x_{\min}$ value is simultaneously selected to minimize the Kolmogorov-Smirnov distance between the empirical and estimated CDFs in $[x_{\min},\infty)$.

\textbf{Degree-replicating weight sequence.}

In general, the empirical degree distributions in our social network dataset only follow a power-law distribution in their tails, hence classical GIRGs (whose weights follow a power law over the whole range $[1,\infty)$) will never perfectly replicate graph statistics closely tied to the degree distribution. Motivated by this, \cite{blaes:fried:katz:krohm:strie} actually uses the degree sequence as weight sequence for the Chung-Lu model. This however fails to make a fair comparison with power-law weighted GIRGs. Therefore we fit two versions of Chung-Lu graphs and GIRGs, one using a degree-replicating weight sequence and one with a power-law weight sequence. 

\subsection{Estimating $c$ and $\alpha$}

The estimation of $c \in (0, \infty)$ and $\alpha \in (1, \infty)$ in the edge probability formula~\eqref{eq:girg-connection} is done jointly. The long-range parameter $\alpha$ monotonically influences the expected local clustering coefficient, which has very low variance for large $n$. Similarly, $c$ monotonically influences the average degree. Since they also cross-influence, we iteratively approximate $\alpha$, then $c$, then $\alpha$ again and so on until the sequences converge. Each estimation is done using binary search, where each evaluation of a parameter pair $(\alpha, c)$ involves generating a GIRG with this parametrization in order to inspect its local clustering coefficient or average degree.

For T-GIRGs, an analytical formula for the expected average degree (for given $c$ and $\alpha$) allows skipping the generation and inspection step in each estimation of $c$.

\subsection{Sampling C-GIRGs from T-GIRGs}

Bläsius et al.\ implement in \cite{blaes:fried:katz:mey:pen:weya} a linear-time sampling algorithm for generating T-GIRGs. As we will demonstrate in section~\ref{sec:experiments}, C-GIRGs often yield more realistic synthetic graphs - Bläsius and Fischbeck already observed that the choice of torus or cube topology can make a substantial difference for the diameter~\cite{blaes:fisch}. Hence, it is useful to be able to sample C-GIRGs efficiently. This can be achieved through an easy coupling, which can bootstrap any black-box T-GIRG sampling algorithm, provided that the sequence of node weights and locations is available (in addition to the graph edges), at a cost linear in the number of edges (which is $\Theta(n)$ for GIRGs) . As the connection probabilities in T-GIRGs stochastically dominate those in C-GIRGs, we can inspect each edge in the generated T-GIRG and determine with a Bernoulli trial with bias $\frac{p_{uv}^\mathrm{C}}{p_{uv}^{\mathrm{T}}}$ whether to keep or delete the respective edge in the C-GIRG, see Algorithm~\ref{alg:cube_coupling}.

\begin{algorithm}
\begin{algorithmic}
    \caption{Generating a C-GIRG from a T-GIRG via coupling}\label{alg:cube_coupling}
    \Require $n$, $d$, $c$, $\tau$, $\alpha$
    \State $\left ( G=(V,E),\; \{x_u\}_{u \in V}, \{w_u\}_{u \in V} \right )\; \gets \text{T-GIRG}(n,d,c,\tau, \alpha)$
    \For{$uv \in E$}
        \State
            $p_{uv}^{\mathrm{T}} = \min \{1, c \left (
            \frac{w_u w_v / W}{Vol(||x_u - x_v||_{\mathrm{T}})} \right )^\alpha \}
            ;\;\;\;
            p_{uv}^{\mathrm{C}} = \min \{1, c \left (
            \frac{w_u w_v / W}{Vol(||x_u - x_v||_{\mathrm{C}})} \right )^\alpha \}
            ;\;\;\;
            p \sim U([0, 1])$
        \If{$p > \frac{p_{uv}^\mathrm{C}}{p_{uv}^{\mathrm{T}}}$}
            \State delete edge $uv$ from $E$
        \EndIf
    \EndFor
    \State \Return $(V,E)$
\end{algorithmic}
\end{algorithm}

\subsection{Sampling Non-Metric GIRGs}

The sampling algorithm for GIRGs introduced in~\cite{bring:keusch:leng:girgs} depends linearly on the number of nodes but scales exponentially in the dimension. In practice, this makes the generation of even moderately large graphs with $n \geq 10'000$ nodes heavily computationally constrained already for dimensions $d=4$ and higher.
To mitigate the problem, we propose an algorithm that allows sampling of an MCD-GIRG in time $O(dn)$ by combining $d$ separately sampled one-dimensional GIRGs (1d-GIRGs) sharing the same node weight sequence; each 1d-GIRG is individually generated using the $O(n)$ algorithm described in~\cite{bring:keusch:leng:girgs}.

For illustration's sake, consider the three-dimensional case.
Let $u,v \in V$ be two vertices with weights $w_u$, $w_v$ and positions $x_u=(x_u^1, x_u^2,x_u^3)$, $x_v=(x_v^1, x_v^2,x_v^3)$, and let $r_1\coloneqq|x_u^1-x_v^1|,r_2\coloneqq|x_u^2-x_v^2|,r_3\coloneqq|x_u^3-x_v^3|$ be the coordinate-wise distances between $u$ and $v$.
The edge probability $p_{uv}$ defined in equation~\eqref{eq:girg-connection} can be viewed as the maximum of $\{p_1, p_2, p_3\}$, where $p_i\coloneqq \min \{1, c \big(\tfrac{w_u w_v / W}{Vol_{\mathrm{MCD}}(r_i)} \big)^\alpha\}$ (with $Vol_{\mathrm{MCD}}$ being the volume function corresponding to the MCD geometry on $[-\tfrac{1}{2},\tfrac{1}{2}]^d$). As $r \rightarrow 0$, we actually have $Vol_{\mathrm{MCD}}(r) = \Theta(r)$ for all dimensions $d$. Therefore, up to constant factors, $p_i$ is equal to the connection probability of vertices $u$ and $v$ in a 1d-GIRG where the positions of $u$ and $v$ are the $i$th coordinate of their position in the torus (and their weights are the same, i.e.\ $w_u, w_v$). This gives rise to Algorithm~\ref{alg:mcd_1d_combination_algo}, which generates a $d$-dimensional MCD-GIRG from $d$ one-dimensional GIRGs (note that in one dimension there is only one possible choice for the distance function, see Definition~\ref{def:boolean-distance-function}).  

The algorithm first generates $d$ different 1d-GIRGs with the same weight sequence, and then determines for each edge in which dimension the distance between its endpoints is the smallest. Both parts have a runtime of $O(dn)$, so the whole algorithm indeed runs in time $O(dn)$.

\begin{algorithm}
\begin{algorithmic}
    \caption{Generating a $d$-dimensional MCD-GIRG from $d$ one-dimensional GIRGs}
    \label{alg:mcd_1d_combination_algo}
    \Require $n$, $d$, $c$, $\tau$, $\alpha$

    \State $V \gets [n];\; E \gets \{\}$
    \State $(w_v)_{v\in V} \stackrel{i.i.d.}{\sim} \mathrm{powerlaw}(\tau)$
    \For{$i \in [d]$}
        \State $\left ( G_i = (V_i, E_i),\; (r_{uv}^{(i)})_{uv \in E_i} \right ) \; \gets \text{1d-GIRG}(n,c,\alpha, (w_v)_{v\in V})$
    \EndFor
    \For{$i \in [d]$}
        \For{$uv \in E_i$}
            \If{$i = \argmin_i r^{(i)}_{uv}$}
                \State $E \gets E \cup \{uv\}$
            \EndIf
        \EndFor
    \EndFor
    \State \Return $G=(V,E)$
\end{algorithmic}
\end{algorithm}

\section{Experiments}\label{sec:experiments}

In this section we report on our experiments\footnote{Our code is available on request.}. We fit 104 Facebook graphs whose sizes range from $n=762$ to $n=35'111$ nodes, retrieved from the Network Repository~\cite{rossi:ahmed}. We sampled max-norm T-GIRGs of all dimensions $d\in\{1,\ldots,7\}$ and MCD-GIRGs (with the torus topology) of dimensions $d\in\{2,3,4,5\}$  (note that for $d=1$ there is no difference between the max-norm and the Minimum-Component-Distance). We also considered a subset of possible Boolean distance functions in dimensions 3 and 4 to generate some other non-metric GIRGs (see Table~\ref{table:results_table} for details).
For C-GIRGs, computational constraints only allowed us to consider dimensions $d\in\{1,2,3\}$ for power-law distributed weights (which requires an estimation of the parameter $\tau$), and dimensions $d\in\{1,2,3,4,5\}$ for degree-replicating weights.

The experiments were run on a High Performance Computing cluster with Intel(R) Xeon(R) CPU E3-1284L v4 @ 2.90GHz CPUs. The main fitting, generating, and feature extracting experiments with the whole range of generative graph models were done with 12 CPUs and 24 GB of RAM, taking between 1 and 2 days to complete.

The feature combinations presented in Table~\ref{table:results_table} are a subset of those presented in~\cite{blaes:fried:katz:krohm:strie} (there the feature combinations were filtered  using a process of numerical, variation, and correlation cleaning). 

We further removed feature combinations that added little extra information (e.g.\ supersets of the feature set "n, m, diam", since these three features are already enough to detect the synthetic networks almost perfectly, and adding extra features to the set only made the misclassification rate even smaller).

\begin{sidewaystable}[ht]

    \centering
    \begin{tabular}{|c|c||c|c|}
        \hline
        1ccu & degree-replicating C-GIRG ($d=1$)
        & n & number of vertices
        \\
        1-23 & $|x_1| \wedge (|x_2| \vee |x_3|)$ non-metric GIRG
        & m & number of edges
        \\
        2m & $|x_1| \wedge |x_2|$ MCD-GIRG ($d=2$)
        & tau & exponent of the power-law degree distribution
        \\
        3cu & C-GIRG ($d=3$)
        & diam & graph diameter 
        \\
        7d & T-GIRG ($d=7$)
        & eff-diam & effective graph diameter 
        \\
        CL & Chung-Lu graph
        & k-core & node $k$-core number 
        \\
        CL-c & degree-replicating Chung-Lu graph
        & LCC & node local clustering coefficient
        \\
        BA & Barab{\'a}si-Albert graph
        & Katz & node Katz centrality 
        \\
        ER & Erd\H{o}s-Rényi graph
        & betw & node betweenness centrality
        \\
        hyper & hyperbolic random graph
        & close & node closeness centrality 
        \\
        \hline
    \end{tabular}
    \caption{Generative graph model abbreviations (cf.\ Section~\ref{sec:graph-models}) and feature name abbreviations (cf.\ Section~\ref{sec:framework}) used in our results table (Table~\ref{table:results_table} below). All the node-specific features refer to the node averages.}
    \label{table:abbreviations}


 \vspace{3\baselineskip}
    \centering
    \includegraphics[width=\textwidth]{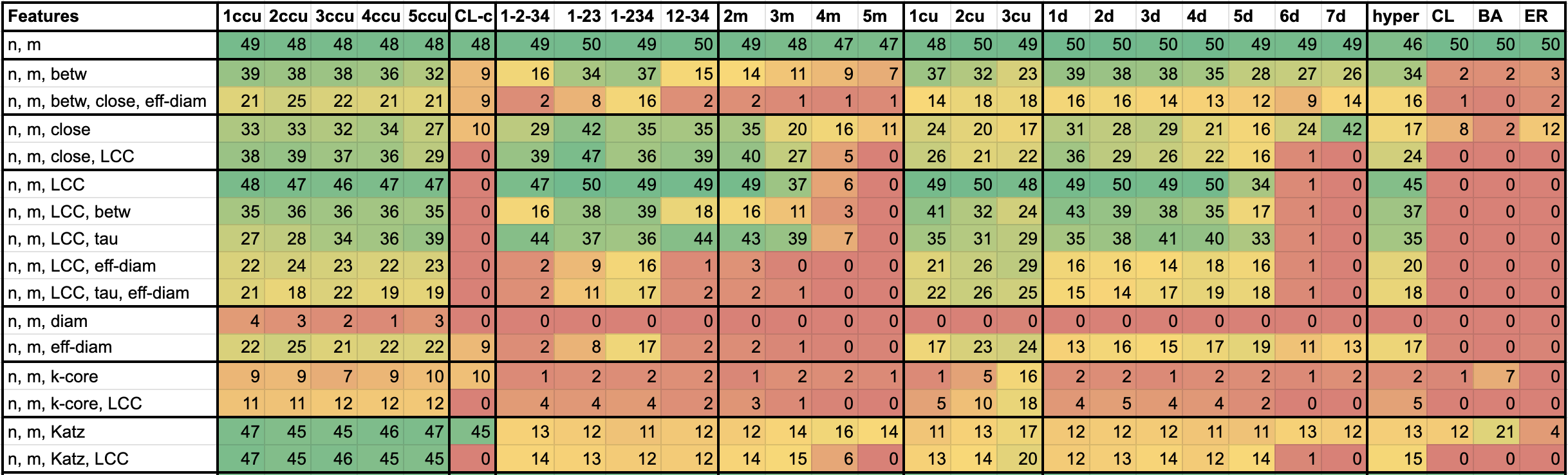}
    \caption{Misclassification rates for different generative graph models and feature sets, see Table~\ref{table:abbreviations} for the meaning of the abbreviations.}
    \label{table:results_table}
\end{sidewaystable}

\subsection{Discussion of the Results}
The key finding can be summarized as follows: GIRGs perform similarly to other generative graph models on non-geometric features, and perform significantly better on geometric features such as closeness and betweenness centrality as well as the local clustering coefficient. Among the previously examined generative graphs models, the only one that is not reliably classified as synthetic is the hyperbolic random graph model~\cite{blaes:fried:katz:krohm:strie}, which is a special case of the GIRG model. Among the different GIRG models, the degree-replicating GIRGs exhibit the best overall performance, in particular when Katz centrality is taken into account. Slightly surprisingly, the degree-replicating weight sequence, which can be seen as overfitting the degree sequence, does not seem to have severe adverse effects on geometric features. This might indicate that GIRGs are in fact a viable approach to modeling more general degree sequences when relaxing the (power-law) assumptions on the weight distribution, which is an interesting research question in its own right, though beyond the scope of the present study. C-GIRGs perform better than T-GIRGs, in particular when diameter or effective diameter are used as features.
On the other hand, models with an underlying outer-min Boolean distance function (including MCD-GIRGs) seem to perform best when closeness centrality is taken into account. Regarding the dimensionality of the considered models, there is no clear monotone relationship between misclassification rate and dimension. One may have conjectured that, at least for relatively small dimensions, the fit improves with increasing dimension (since this yields more free parameters), but often the misclassification rate is already at or near the peak for one- or two-dimensional models. It may well be that this is further evidence for the inherent low-dimensionality of real-world networks. Friedrich et al.\ have recently shown that, as the dimension goes to infinity, the GIRG model converges to the Chung-Lu model~\cite{fried:goeb:clique} - which has mostly zero misclassification rate for the geometric features. It is possible that this behavior is already visible at low dimensions and could explain the decreasing misclassification rates for some of the geometric features.
For example, it is known that the clustering coefficient in GIRGs decays exponentially with the dimension~\cite{fried:goeb:stat}, hence it is inherently difficult for such models to produce a large clustering coefficient in high dimensions.
One further shortcoming of current GIRG variants seems to be the tightly concentrated graph statistics, such as the clustering coefficient - a characteristic where practice and theory are at odds, since real-world networks exhibit a much larger variance for most node-specific features. Another example of a significant difference between synthetic graphs and real-world networks is the diameter, see Table~\ref{table:results_table}. It turns out that the diameters of all generated graphs are small (taking values between 3 and 5) compared to real-world networks (where the diameters range from 6 to 12). These shortcomings demonstrate the need for extended models which allow for better approximations of these features.

\section{Conclusion}

We extend the framework introduced in~\cite{blaes:fried:katz:krohm:strie} by Bläsius et al.\ to cover a large class of Geometric Inhomogeneous Random Graphs - both metric and non-metric - finding that they are effective at approximating geometric features of real-world networks. To enable this extension, we devise a joint estimation technique for the connection probability constant $c$ and the long-range parameter $\alpha$, as well as two fast GIRG-sampling algorithms building on the well-known algorithm introduced by Bringmann, Lengler and Keusch in~\cite{bring:keusch:leng:girgs}. Several exciting avenues of follow-up research extend from here. As more high-quality real-world network data becomes available and new types of networks emerge, it would interesting to see how well these novel networks can be modeled, by GIRGs specifically and in the framework at large. In particular, we would like to see findings for biological and technological networks. A natural next step would be to include disconnected networks - whereby features such as diameter and average distance could be analyzed component by component, or alternatively restricted to the largest connected component of each network. A different and particularly promising perspective compares algorithmic performance on different types of network models and real-world networks. This is not limited to classical algorithms such as Breadth-First-Search and Routing Algorithms, but could extend to large classes of distributed algorithms and even other stochastic processes such as epidemic models, rumor spreading and opinion forming. We believe that here the investigation of further non-metric GIRGs, as well as so called assortative and disassortative models (i.e. models exhibiting degree-degree correlations), will provide valuable insight. In general, the framework extends easily to additional generative graph models, such as (degree-corrected) stochastic block models and (spatial) preferential attachment models, and comparing their performances would be
a promising next step. Beyond the binary classifier put forth in~\cite{blaes:fried:katz:krohm:strie}, a wealth of alternative architectures exist, namely approaches such as Graph Neural Networks and Large Multimodal Models.  Finally, once our models approximate the networks sufficiently in terms of single-value statistics, their distributions should be investigated.

\paragraph{Acknowledgments.} We thank Johannes Lengler for valuable discussions and feedback. M.K. and U.S. gratefully acknowledge support by the Swiss National Science Foundation, grant number 200021\_192079.

%
%

\end{document}